# New Phases and Dissociation-Recombination of Hydrogen Deuteride to 3.4 Megabar

Ranga P. Dias, Ori Noked, and Isaac F. Silvera

Lyman Laboratory of Physics, Harvard University, Cambridge MA, 02138

We present infrared absorption studies of solid hydrogen deuteride to pressures as high as 3.4 megabar in a diamond anvil cell and temperatures in the range 5 to 295 K. Above 198 GPa the sample transforms to a mixture of $HD$, $H_2$ and $D_2$, interpreted as a process of dissociation and recombination. Three new phases-lines are observed, two of which differ remarkably from those of the high-pressure homonuclear species, but none are metallic. The time-dependent spectral changes are analyzed to determine the molecular concentrations as a function of time; the nucleon exchange achieves steady state concentrations in ~20 hours at ~200 GPa.

In 1935 Wigner and Huntington predicted that hydrogen would transform to an atomic metallic solid at high density [1]. There is a decades long effort to metallize the hydrogens ($H_2$, $D_2$, and HD) in the solid state at high pressure and low temperature or moderate pressures and high temperatures. We have studied HD to very high pressures. With increasing pressure we first observe the line separating phases I-III; then when entering a new phase at 198 GPa, $HD$ transforms to $HD$, $H_2$, and $D_2$. We propose that the nucleon exchange is due to a process of dissociation followed by recombination that we call **DISREC** (see discussion ahead). At still higher pressures and low temperatures, yet another phase is entered. In contrast to pure isotopic systems, the new phase lines are almost vertical and appear to intersect the T=0 Kelvin axis. None of the phases are metallic.

The hydrogens form remarkable solids whose properties are dominated by quantum effects: the ortho-para isomers due to the Pauli principle and large zero-point energies/motions in the crystal lattice. At low pressure, ortho and para, fundamentally distinguished by their nuclear spin states, have distinct structures and properties. The ortho-para concentrations are metastable with slow conversion to equilibrium [2]. At megabar pressures there are predictions that hydrogen may be a liquid at T=0 K or low temperature due to the zero-point energy [3, 4]. HD is a special case because the nucleons are distinguishable. The free molecules have a small permanent electric-dipole moment ($5.85 \times 10^{-4}$ D) [5]. Our ultimate goal is to extend the phase diagram to higher static pressures and low or modest T for any of the isotopes to achieve the metallic phase. A first-order phase transition to liquid metallic hydrogen (**MH**) at static pressures and high temperatures has recently been observed [6], and liquid metals of hydrogen and deuterium has been observed in high-temperature shock experiments [7, 8]. There have been a number of very high static pressure studies of the solid homonuclear species at lower temperatures, but none have achieved metallization [9-14].

To understand *HD* in the context of the hydrogens we first discuss their known phase diagrams. Ideally the structures of phases are determined by x-ray or neutron diffraction techniques, but these are very challenging measurements for the hydrogens. In practice their phase lines have been determined by optical methods, studying Raman or IR active modes. At a phase transition, modes disappear, appear, or undergo discrete shifts as the symmetry of the lattice changes, and these changes pinpoint the phase lines. Samples with mixed ortho-para concentrations lack translational symmetry but still demonstrate similar phase transitions as determined by optical measurements. The loss of translational symmetry is more distinct for isotopic mixtures due to the different masses, yet we observe clear evidence of phase transitions. We believe that these transitions are structural, for example, transitions to structures having graphene like planes with translational disorder within the planes.

At high pressures, solid $H_2$, $D_2$ and *HD* have three distinct common phases shown in Fig. 1. There, we show the known phase diagram of the hydrogens as well as our new data, which we shall discuss later. Since the phase diagrams of hydrogen and deuterium are similar, for comparison to HD we shall focus on one of these, $D_2$. For <u>pure</u> solid ortho-$D_2$, a transition from the low pressure (LP) hexagonal phase to an orientationally ordered (OO) structure called the BSP (broken symmetry phase) takes place at 28 GPa [15]; a second OO transition occurs at ~150 GPa [16] to the D-A phase. For <u>mixed</u> ortho-para concentrations, which lack translational symmetry, the LP, BSP, and D-A are called phases I, II, and III. Recently, at higher pressures

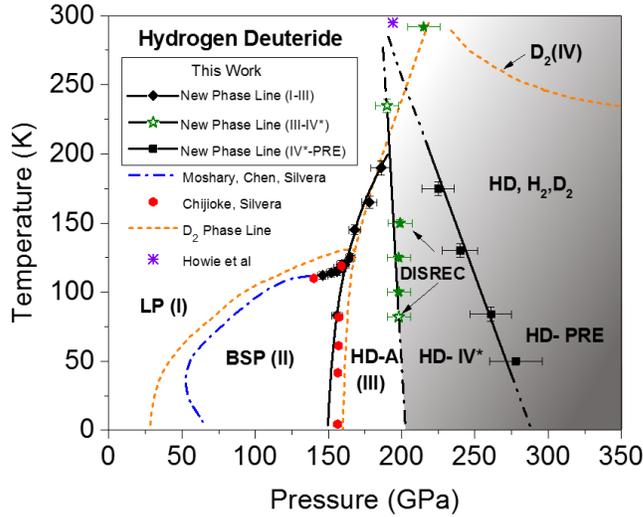

Fig. 1. The phase diagram of *HD* extending the earlier measurements beyond the triple-point to establish the I-III line. Solid lines indicate new phase-lines of mixtures of isotopes. For comparison we show the phase diagram of $D_2$. For the III to HD-IV* line we only traversed the phase line at P~198 GPa and T=82 and 240 K, indicated by open stars, while

(~230 GPa) and around room temperature, an insulating phase named IV has been observed for $H_2$ and $D_2$ [11, 12, 17].

A number of years ago Brown and Daniels [18] showed that when a mixture of $H_2$ and $D_2$ is pressurized (under 100 GPa) at room temperature, a process takes place to form HD, leaving an $H_2$, $D_2$, HD mixture (this process has recently been confirmed with higher pressure studies by Howie et al up to almost 300 GPa [19]). Subsequently, Moshary, Chen, and Silvera [20] studied a sample of HD, showing that *$2HD \rightarrow H_2 + D_2$* does not take place to at least 110 GPa at 77 K, while Chijioke and Silvera [21] saw no evidence of sample changes to 159 GPa and 160 K. In this letter we report on a study of HD to a pressure of ~340 GPa. No changes of HD were observed after 14 hours at 184 GPa and 82 K. Above 198 GPa and temperatures ranging from 5 to 295 K, we observe the formation of $H_2$ and $D_2$, while HD remains a component. We extend the phase diagram for HD, studied earlier [21, 22], to our highest pressures. The lattice is not metallic at the pressures studied and we conclude that the process *$2HD \leftrightarrow H_2 + D_2$* can take place at ultra-high pressure.

The samples of HD gas (our virgin sample of *HD* had ~4% $H_2$ impurity) were cryogenically loaded in diamond anvil cell (DAC). The DAC was mounted in a cryostat similar to an apparatus described elsewhere [23]. Conic shaped type II diamonds with 50-micron diameter culets were coated with alumina that acts as a hydrogen diffusion barrier; a rhenium gasket was used. IR spectra were obtained in the temperature range 5-295 K with a Nicolet FTIR interferometer that uses an internal thermal light source. The experiment lasted 6 weeks and terminated with failure of the diamond anvils. At the highest pressures the diameter of the sample was about 14 microns (see photo in Supplementary Information, SI); interferograms were averaged for about 35 minutes to obtain adequate signal-to-noise (see SI). Although the sample contained a ruby particle for pressure determination, our studies extended far beyond the calibrated ruby scale [24]. As is common for ultra high-pressure experiments on the hydrogens, we used frequencies of the IR vibron modes for pressure determination [25, 26] (see SI).

Spectra of the sample were obtained as a function of pressure and temperature. In Fig. 2 we show fits to the IR absorption spectra (also see SI). Remarkably, the IR spectral lines that we observe from the three species remain rather sharp and resemble spectra in the pure isotopic samples, in spite of the conceptual picture that any molecule is in a cluster of neighbors of mixed isotopic composition. This is probably because the intermolecular interactions average out for mixed samples and impurity modes (impurity isotopes in an isotopically pure sample) are experimentally known to follow the IR mode pressure dependence, as discussed in ref. [20]. Figure 2a is at fixed temperature (82 K). At 198 GPa we see the appearance of a $D_2$ vibron line (after ~11 hours), splitting of the $HD$ vibron line, and growth of the $H_2$ vibron line intensity; all spectral lines clearly show a splitting at 261 GPa. Figures 2b,c show the crossing of phase lines at fixed pressures and varying temperatures. The splittings, shifts, and changes for all of our measurements are summarized in Fig. 3, which shows the peak frequencies of the IR lines as a function of pressure at 82 K.

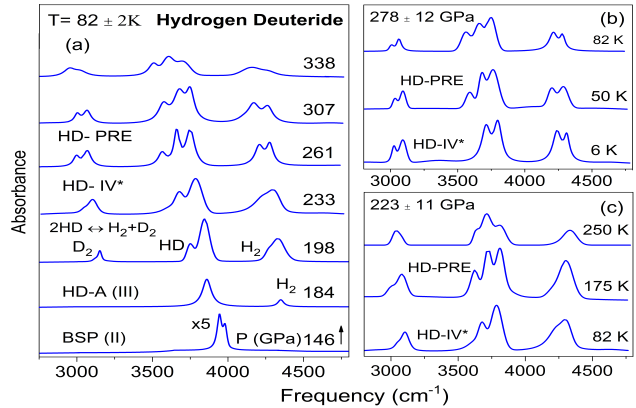

Fig. 2. Left: infrared spectra as a function of pressure at a temperature of 82 K. Right: IR spectra as a function of temperature for the indicated pressures. At 82 K and 146 GPa the impurity hydrogen vibron line is too weak to be observed in phase II. With increasing pressure one sees the onset of deuterium lines and growth of hydrogen lines, as well as splitting of the spectral lines.

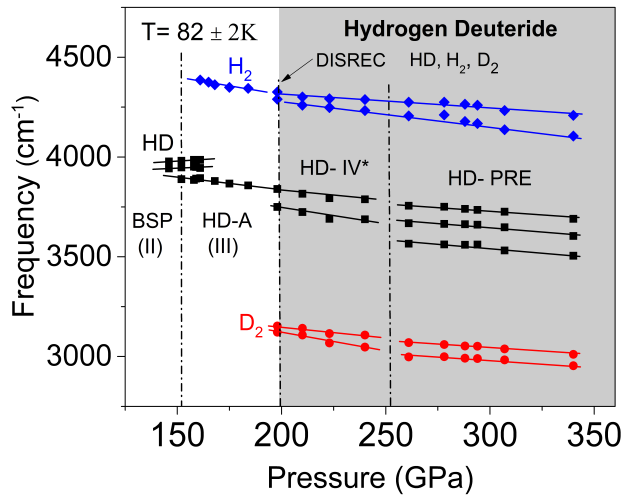

Fig. 3. Peak frequencies of the IR modes as a function of pressure, indicating phase transitions where splittings or new modes appear. The shaded area is where a mixed isotope sample, due to DISREC was observed.

Figure 1 shows the P,T data points for the phases of HD. First of all, our data confirms the triple point found in ref. [21] and we determine the phase-line between phase I and III. We observe two additional phase-lines, very different from the phase-line for pure deuterium or hydrogen. We call the lower-pressure phase HD-IV*; the higher-pressure phase may precede the transition to metallic hydrogen deuteride and we have named it *HD*-PRE. Both phases have negative slopes with regard to pressure and extrapolate to 200 and 290 GPa at the T=0 K axis. These phase lines, that are almost vertical, are quite different from the behavior in the pure isotopic samples, with phase IV.

Simulations have shown that there are several structures, close in energy, that are composed of hexagonal graphene-like planes for the pure isotope solids [27-32]. We do not have an explanation for the difference between the phase lines for pure or mixed isotope samples exhibited in our data. We comment that in the pure homonuclear species the transition is believed to be entropy driven [30] which may be mitigated by a mixed molecular sample. Raman spectroscopy supports a structure with alternating hexagonal layers in the pure isotopic solids. Two distinct Raman vibrons were observed, corresponding to the two planes. We note that only the structures of the lowest-pressure phases are known [2] and the structures of phases II-IV are at best calculated and compared to spectroscopic observations.

Phase IV of $H_2$ and $D_2$ has been interpreted by Howie et al [12] as being composed of distinct hexagonal layers, called a mixed molecular-atomic phase with molecular and atomic layers, however our measurements, as well as a simulation [30], do not support this assertion. The existence of the two distinct vibron modes arising from two distinct sets of planes is incompatible with the existence of atomic layers; planes of atoms do not have vibron modes. If the model proposed by Howie et al [12] were correct, that alternate layers are atomic, then when phase IV is entered, half of the molecules dissociate and this should have a large impact on the intensity of the vibron modes.

For our high-pressure observations we believe that the mechanism to form hydrogen and deuterium molecules is a process involving fluctuating dissociation, followed by recombination with atoms from similarly dissociated neighbors that we call DISREC. This is theoretically supported by the simulations of Liu and Ma (see their SI) [31] who showed that of the alternate hexagonal layers, layer I contained orientationally disordered molecules and layer II orientationally ordered molecules (Ref. [30] names these B and G-layers, respectively). Layer I was stable against nucleon transfer with increasing pressure, while the molecules in layer II (which have expanded bonds) could dissociate and recombine. We believe that this process occurs in HD. For random processes DISREC would lead to a 25-25-50% concentration of $H_2$, $D_2$, and HD, respectively.

HD can be used as a diagnostic tool to understand very high-pressure structures. Since we observe that HD molecules are stable until phase *HD-IV\** is entered, then one should see a growth of the $H_2$ and $D_2$ concentration according to the DISREC model. In transitioning from phase III to *HD-IV\** we observe the growth of $H_2$ and $D_2$ IR modes. When the molecular concentrations arrive at equilibrium, the sample composition is ~56% *HD*, ~26% $H_2$ and ~18% $D_2$ (our *HD* sample starts with 4% $H_2$ impurity) determined from the integrated intensities of the IR modes. This implies that for pressures above ~198 GPa, the entire sample undergoes DISREC as the composition is ~ *HD + 1/2 ($H_2$+$D_2$)*, when our final concentrations are corrected for the initial $H_2$ impurity.

Howie et al [19] studied $H_2$, $D_2$, HD <u>mixtures</u> at room temperature and observed splitting of the Raman vibron lines above about 195 GPa, suggesting a transformation to phase IV (called IV') correlated to the vibron frequencies arising from different layer types. However, they did not study the temperature dependence of the phase line. A linear extrapolation of the HD-IV*-HD-PRE phase lines (our data) to room temperature (Fig. 1) shows that with increasing pressure at room temperature they may have been in the HD-PRE phase, rather than the phase they identified as similar to IV and named HD-IV'.

In our mixed sample the HD IR vibron line shown in Fig. 3 splits into three lines with the entry into the HD-PRE phase, while $H_2$ and $D_2$ vibron lines do not split, which is unexpected. We have compared our IR data with the Raman data of Howie et al on mixed samples [19]. They suggest that the transformation at 198 GPa is one involving double hexagonal layers. Apparently they used the pressure dependence of the Raman vibrons from Phase IV of the pure species for layers I (B) and II (G) to identify their structure. In the SI we re-plot some of their data for a 50-50 $H_2$-$D_2$ mixture that after transformation should be quite similar to our composition. For the pure species their B modes are approximately constant with increasing pressure while the G modes have a steep negative slope. This is the case for the modes that they associate with deuterium and hydrogen deuteride, but this is not the behavior for the hydrogen G mode, implying a problem with the mode identification. This is a complex system that should be reanalyzed to investigate possible structures and associated Raman and IR active modes.

We have used the integrated intensities of the IR lines to determine the concentrations of the various species as a function of time at a pressure of 198 ± 8 GPa and temperature of 82 ± 2 K. Observation of the sample for 14 hours at P=184 GPa, T=82 K showed no DISREC, i.e. the *HD* concentration was stable. The change of concentration vs. time for HD-IV* is plotted in Fig. 4. To describe the spectral kinetics data (see Fig. 4), we used Avrami's equation [33-36] $(x(t)-x_e)/(x_0-x_e)=\exp(-bt^n)$, where $x$ is the concentration of a species at time $t$, $x_o$ and $x_e$ are initial and equilibrium values, and $b$ and $n$ are the parameters relating to the rate of reaction and the processes that characterize the reaction (the Avrami exponent). The reaction rate $k$ is equal to $b^{1/n}$. We obtained the best fit with the Avrami parameters (see inset in Fig. 4) of $n_{HD}=2\pm0.02$, $n_{H_2}=1.75\pm0.04$, and $n_{D_2}=1.7\pm0.08$, which indicates that the reaction is in the nucleation dominated regime [37]. That is, the composition at the point sampled is influenced by the nucleation growth of the different molecules in that region, so the process is homogenous.

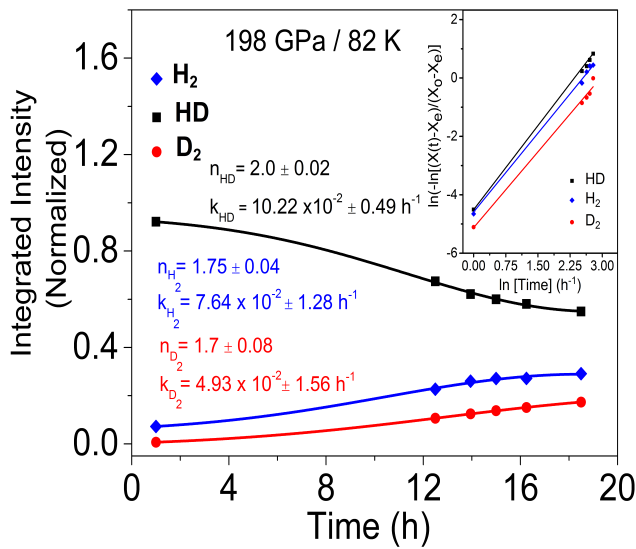

Fig. 4. The normalized IR integrated intensity changes of hydrogen isotopes plotted as a function of time, showing the kinetics associated with the DISREC at 198 GPa and 82K. These time-dependent intensity changes are fitted to Avrami's equation (see the inset) to yield the rate constants and Avrami exponent marked on the plots.

In summary, we have studied hydrogen deuteride up to ~340 GPa and compared its behavior with that of pure molecular hydrogen and deuterium. Three phase lines were determined: HD-IV*, *HD*-PRE, and the I-III line. *HD* transforms to isotopic mixtures for pressures around 200 GPa and higher, interpreted as a dissociation and recombination process. When HD-IV* is entered, DISCREC commences. Our view is that in this region the structure is layered; a negligible density of atomic species exists and all atoms are bound as molecules, probably with bond lengths that depend on the planes they occupy. Atoms can exist due to bond-breaking fluctuations, but they rapidly recombine, so that the expectation value of free atoms in the lattice is very small. It is also unclear why new high-pressure phases reported here are so different from those of the homonuclear species. In the goal of metallization of the hydrogens, HD remains an important candidate. Low density HD is also of special astronomical interest with the recent detection of HD in interstellar clouds [38, 39]. Deuterium is believed to have been created primordially in the Big Bang nucleo-synthesis. Determination of deuterium abundance, relative to hydrogen in the atmospheres of planetary systems and nebulae provides important cosmological constraints on planetary and stellar formation [40]. In this regard, the dissociation and recombination process is quite important and abundances must also consider the HD concentrations.

We thank M. Zaghoo, A. Salamat, G. Ackland, G. Borstad, and R. Husband for discussions of the properties of the hydrogens. The NSF, grant DMR-1308641 and the DOE Stockpile Stewardship Academic Alliance Program, grant DE-FG52-10NA29656 supported this research. Preparation of diamond surfaces was performed in part at the Center for Nanoscale Systems (CNS), a member of the National Nanotechnology Infrastructure Network (NNIN), which is supported by the National Science Foundation under NSF award no. ECS-0335765. CNS is part of Harvard University.